\documentclass[a4paper]{article}

\usepackage{INTERSPEECH2020}
\usepackage{amsmath,graphicx}
\usepackage{booktabs}
\usepackage{xcolor}
\usepackage{multirow}
\usepackage{caption}
\usepackage{subcaption}
\usepackage{booktabs}
\usepackage{afterpage} 
\usepackage{float}
\definecolor{green_print}{RGB}{94,60,153}
\definecolor{red_print}{RGB}{230,97,1}
\title{Speaker discrimination in humans and machines: Effects of speaking style variability}
\name{Amber Afshan$^1$, Jody Kreiman$^2$, and Abeer Alwan$^1$\thanks{This study was supported in part by NSF}}
\address{
  $^1$Dept. of Electrical and Computer Engineering, University of California, Los Angeles, USA\\
  $^2$Depts. of Head and Neck Surgery and Linguistics, University of California, Los Angeles, USA}
\email{amberafshan@g.ucla.edu, jkreiman@g.ucla.edu, alwan@g.ucla.edu}

\begin{document}

\maketitle
\begin{abstract}
Does speaking style variation affect humans' ability to distinguish individuals from their voices? How do humans compare with automatic systems designed to discriminate between voices? In this paper, we attempt to answer these questions by comparing human and machine speaker discrimination performance for read speech versus casual conversations. Thirty listeners were asked to perform a same versus different speaker task. Their performance was compared to a state-of-the-art x-vector/PLDA-based automatic speaker verification system. Results showed that both humans and machines performed better with style-matched stimuli, and human performance was better when listeners were native speakers of American English. Native listeners performed better than machines in the style-matched conditions (EERs of 6.96\% versus 14.35\% for read speech, and 15.12\% versus 19.87\%, for conversations), but for style-mismatched conditions, there was no significant difference between native listeners and machines. In all conditions, fusing human responses with machine results showed improvements compared to each alone, suggesting that humans and machines have different approaches to speaker discrimination tasks. Differences in the approaches were further confirmed by examining results for individual speakers which showed that the perception of distinct and confused speakers differed between human listeners and machines.
\end{abstract}

\section{\label{sec:intro}Introduction}
Speaking style variations are prevalent in everyday life, changing as we move from talking to a friend to reading aloud, from public speaking to talking to an infant. Regardless of these variations, humans are often able to recognize a familiar voice after hearing it for a few seconds~\cite{wenndt_machine_2012}. Previous research suggests that for humans, recognizing familiar talkers entails matching a sample to stored voice templates, whereas recognizing unfamiliar talkers is a much more involved process requiring acoustic feature comparisons~\cite{van_lancker_voice_1987}. In this work, we are interested in understanding the effects of speaking style variations on the abilities of humans to distinguish between unfamiliar voices from short duration ($\sim$3~s), text-independent utterances. We are also interested in comparing human performance against state-of-the-art automatic systems.

A speaker rarely says an utterance twice in the exact same manner. This variability maybe intentional--for example, to hide one's identity or to communicate irony--or it may be introduced without conscious intention due to changes in emotion, social context, or physiological state~\cite{kreiman_foundations_2011}. These voice variations have been extensively studied in the forensic literature~\cite{saslove_long-term_1980, blatchford_idenfication_2006, gonzalez_hautamaki_limits_2019, gonzalez_hautamaki_automatic_2015}. For example, style variability confuses ear witnesses hearing a criminal shouting vs reading aloud during a voice lineup~\cite{jessen_forensic_2008}. Human and machine speaker discrimination performances have been compared when style changed from read to pet-directed speech, which is characterized by exaggerated prosody~\cite{park_towards_2018}.
In both examples, differences in style were extreme, and little is known about how moderate variations in style, for example between read and conversational speech, affect the relative performance of humans vs machines in speaker discrimination performance.  Evidence from voice sorting tasks indicates that humans vary their perceptual strategies when ``telling people together'' versus ``telling people apart'' \cite{lavan_how_2019, johnson_comparing_2019}, while machines apply the same classification approach in target and non-target trials.  Given that humans outperform machines in some tasks [e.g., ~\cite{park_target_2019}], this suggests that machines can adopt strategies from humans, and humans might do better with machine assistance in certain situations.  Additionally, it has previously been observed that there is a correlation between the linguistic proficiency of the listener and their speaker identification accuracy~\cite{perrachione_learning_2007,schweinberger_speaker_2014}. There is uncertainty, however, whether the effects of linguistic proficiency are consistent across different speaking styles. 

In this study, we employed an unfamiliar \emph{speaker discrimination} task in which the listener decides if two samples are from the same speaker or not. 
Using read and conversational speech stimuli, we compared the discrimination performance and reliability of humans and machines when confronted with moderate differences in speaking style.  We hypothesized that such variations will have different effects on automatic speaker verification (ASV) systems than they do on human performance. 
We also examined the relationship between listeners' native languages and their accuracy in performing speaker discrimination tasks for different speaking styles. 

The paper is organized as follows. Section~\ref{sec:data} describes the databases used. Methods for machine and human experiments are detailed in Section~\ref{sec:method}. Results and discussion are presented in Section~\ref{sec:resndis}, and we conclude the paper with Section~\ref{sec:con}.

\section{\label{sec:data}Databases}
\subsection{The UCLA speaker variability database}
The UCLA Speaker Variability Database~\cite{keating_new_2019,kreiman_relationship_2015} captures commonly occurring variations in speech from 103 female and 105 male speakers. These variations were due to 
phonetic content, speaking style, and affect conditions. Speech was recorded in a sound-attenuated booth at a sampling rate of 22kHz. 
The experiments reported in this study used a subset of recordings from 40 female talkers who were self-reported native speakers of American English (confirmed \textit{post hoc} by two linguists). In order to avoid gender dependent cues, only female talkers were chosen. Additionally, female talkers had discernible prosody changes between speaking styles in comparison to male talkers. Two types of speech samples were selected for each talker. The first included five phonetically-rich Harvard sentences~\cite{noauthor_ieee_1969}, read twice in random order (clear read speech); the second comprised the talker’s side of a 2-minute telephone conversation with a family member or friend (casual conversational speech).

\subsection{NIST SRE and Switchboard databases}
We used the NIST SRE 04, 05, 06, 08 and 10 databases~\cite{przybocki_nist_2004,przybocki_nist_2006,martin_nist_2009} and  the Switchboard  II  corpus, phase  2~\cite{graff_switchboard-2_1999} to train the ASV system. 
These databases provide more than 3,000 hours of speech samples from 3,408 female and 1,832 male talkers, sampled at 8 kHz. Recordings from the UCLA database were downsampled during the ASV experiments to match this rate.

\section{\label{sec:method}Method}

\subsection{\label{ssec:perceptual}Perceptual speaker discrimination}

Perceptual tasks included trials with two different read sentences, trials with two different sentences excerpted from a conversation, and trials with one read sentence and one conversational sentence. In each case, equal numbers of ``same speaker'' (target) and ``different speaker'' (non-target) trials were included, for a total of six kinds of trials. The stimuli were all $\sim$3~sec long, and long silences before and after the sentences were removed.

Distinct stimuli were selected for each session so that a listener never heard the same stimulus twice. In the case of read speech, there were only five sentences, so a second recording of one of the sentences was repeated at random in each session. 
Among the conversations, six different snippets were chosen at random. Selections were carefully made to ensure that semantic cues would not bias responses. All non-speech vocalizations (laughing, giggling, sighing) were deleted. 

To minimize fatigue, each listener heard a subset of 24 talkers selected at random from the pool of 40, for a total of 144 trials/listener (6 trial types x 24 talkers). Fifteen normal-hearing subjects heard each subset. Here we report results from two groups of listeners who heard 2 different subsets of the voices. Twenty-four/30 listeners were native English speakers; 22 were female and 8 were male. They ranged in age from 17-21. The non-native English speakers consisted of 3 native speakers of Spanish, 2 of Mandarin, and 1 of Hindi.

Stimuli were randomized prior to each presentation.  During
the listening experiments, each pair of stimuli could be played only once in each presentation order (AB/BA). Listeners were asked to decide if the stimuli represented the same talker or two different talkers. They also reported their confidence in their response on a scale of 0 to 5 (0=wild guess and 5=very confident). Listeners were not aware of the number of talkers included in each experiment, and were encouraged to complete the experiments at their own pace, and take breaks as necessary.  Testing time averaged about 45 minutes.

\subsection{Automatic Speaker Discrimination}
An x-vector~\cite{snyder_x-vectors_2018} /PLDA~\cite{kenny_plda_2013} (probabilistic linear discriminant analysis) based ASV system was used, and the PLDA was adapted~\cite{garcia-romero_unsupervised_2014} using in-domain data with both read and conversational styles to achieve the best possible machine performance. 
In order to ensure a fair comparison between humans and the ASV system, trials matched the stimulus pairs presented to listeners. We followed the Kaldi~\cite{povey_kaldi_2011} SRE16 recipe to train the x-vector embedding extractor. Mel-frequency cepstral coefficients of 23 dimensions calculated using a frame-length of 25ms were used as features for the x-vectors. The system was trained using the SRE and Switchboard databases.

\subsection{\label{ssec:eval}Evaluation Metric}

\subsubsection{\label{sssec:scores}Calculation of scores}
Human responses were unfolded to obtain a similarity score between each stimuli pair. Confidence ratings ($0$ to $5$) were multiplied by the decision (different=$-1$ and same$=1$) to provide continuous scores between $-5$ (highly confident that the voices are different) and $5$ (highly confident that the voices are same). This ensured that the similarity score reflected the decision as well as the confidence of human responses. For ASV systems, the PLDA score acted as the similarity measure. The PLDA score represents the ratio of the likelihood that a given pair of stimuli is from the same speaker to the likelihood that the pair is from two different speakers. 

Calibrated log-likelihood ratios (LLR; L) were obtained from similarity scores using a calibration system based on standard logistic regression~\cite{brummer_bosaris_2011}. The resulting LLRs represent scalar responses by humans and machines.

\subsubsection{\label{sssec:eercllr}Analysis of performance errors}
System-level speaker discrimination performance was evaluated in terms of equal error rates (EER) and the log-likelihood-ratio cost function ($C_\textrm{llr}$) \cite{van_leeuwen_introduction_2007}. While the EER is a widely used measure, it does not measure calibration, the ability to set good decision thresholds. Hence, $C_\textrm{llr}$, an application-independent measure for evaluating soft decisions, was also used. 
It can be interpreted as a measure of loss of information; thus, the lower the $C_\textrm{llr}$, the more the average information per trial (in bits) increases by applying the system (humans or machines). 



To perform calibration and to calculate the evaluation measures, we used the Bosaris toolkit~\cite{brummer_bosaris_2011}. As a result of the limited amount of data, the calibration parameters were trained and applied on the same set of scores.

\subsubsection{\label{ssec:fusion} System Fusion}
In order to better understand the similarities and differences between responses from humans and machines on the speaker discrimination task, logistic regression-based~\cite{brummer_fusion_2007} system fusion was performed on the log-likelihood-ratios from humans and machines using the Bosaris toolkit~\cite{brummer_bosaris_2011}. If humans and machine use different strategies in performing this task, system fusion should outperform either system alone. 
\subsubsection{\label{sssec:spk_lvl}Speaker-Level Analysis}
This section describes speaker-level equivalents of the measures described in the previous sections. The log-likelihood-ratio $L_t$, as outlined in Section~\ref{sssec:scores}, was obtained for each trial $t$. $L_t$ represents the scalar response by the system for the given trial. To compare the scores between target and non-target trials for each speaker, the $L^\textrm{tar}$ for target and $L^\textrm{non}$ for non-target trials were calculated separately. 

$L^\textrm{tar}$ indicates within-speaker variability: a large $L^\textrm{tar}$ means small within-speaker variability (i.e., these target trials are easy). $L^\textrm{tar}$ for a speaker was obtained by averaging the $L_t$ values over the target trials that included that particular speaker. On the other hand, $L^\textrm{non}$ represents between-speaker variability, so a large  $L^\textrm{non}$  value indicates that the speaker did not differ very much from others, so that it is difficult for the system to distinguish her from others. 

A speaker-level aggregation of the log-likelihood-ratio cost function ($C_\textrm{llr}$; section ~\ref{sssec:eercllr}) was also computed, and represents  the confidence the system has when identifying a speaker. Speaker-level $C_\textrm{llr}$ values for target trials ($C_\textrm{llr}^\textrm{tar}$) and non-target trials    ($C_\textrm{llr}^\textrm{non}$) were also computed.

\begin{table*}[t]
\centering
    \caption{Speaker discrimination performance in terms of equal error rates (EER, \%) and log-likelihood-ratio cost function: combined ($C_\textrm{llr}$), target trials $C_\textrm{llr}^\textrm{tar}$ and non-target trials $C_\textrm{llr}^\textrm{non}$.  A comparison of $C_\textrm{llr}^\textrm{tar}$ and $C_\textrm{llr}^\textrm{non}$ is also shown by using purple for the better of the two values in each condition and orange for the worse value.}
    \vspace{-0.5em}

    \label{tab:perf}
\resizebox{0.95\textwidth}{!}{%
\begin{tabular}{cc|c|ccc|c|ccc|c|ccc}
\toprule
\toprule
\multirow{2}{*}{\textbf{Listeners}} &
  \multirow{2}{*}{\textbf{System}} &
  \multicolumn{4}{c|}{\textbf{read-read}} &
  \multicolumn{4}{c|}{\textbf{conversation-conversation}} &
  \multicolumn{4}{c}{\textbf{read-conversation}} \\ \cmidrule{3-14}
 &
   &
  \textbf{EER \%} &
  \textbf{$C_\textrm{llr}$} &
  \textbf{$C_\textrm{llr}^\textrm{tar}$} &
  \textbf{$C_\textrm{llr}^\textrm{non}$} &
  \textbf{EER \% } &
  \textbf{$C_\textrm{llr}$} &
  \textbf{$C_\textrm{llr}^\textrm{tar}$} &
  \textbf{$C_\textrm{llr}^\textrm{non}$} &
  \textbf{EER \%} &
  \textbf{$C_\textrm{llr}$} &
  \textbf{$C_\textrm{llr}^\textrm{tar}$} &
  \textbf{$C_\textrm{llr}^\textrm{non}$} \\ \midrule
\multirow{3}{*}{Native}     & Machines & 14.35 & 0.4421 & {\color{red_print}0.4443} & {\color{green_print}0.4399} & 19.87 & 0.5765 & {\color{red_print}0.5888} & {\color{green_print}0.5642} & 21.78 & 0.6492 & {\color{red_print}0.6501} & {\color{green_print}0.6484} \\ 
                            & Humans    & 6.96  & 0.2642 & {\color{green_print}0.2103} & {\color{red_print}0.3182} & 15.12 & 0.5291 & {\color{green_print}0.5010} & {\color{red_print}0.5573} & 20.68 & 0.6911 & {\color{green_print}0.6900} & {\color{red_print}0.6923} \\ 
                            & Fusion   & 4.92  & 0.1731 & {\color{green_print}0.1447} & {\color{red_print}0.2015} & 11.20 & 0.3790 & {\color{green_print}0.3624} & {\color{red_print}0.3956} & 16.39 & 0.5207 & {\color{red_print}0.5213} & {\color{green_print}0.5202} \\ \midrule
                            
\multirow{3}{1cm}{\centering Non Native} & Machines & 13.95 & 0.4113 & {\color{red_print}0.4147} & {\color{green_print}0.4079} & 19.47 & 0.5682 & {\color{red_print}0.6116} & {\color{green_print}0.5248} & 19.64 & 0.6754 & {\color{red_print}0.6791} & {\color{green_print}0.6718} \\ 

                            & Humans    & 12.39 & 0.4292 & {\color{green_print}0.3836} & {\color{red_print}0.4748} & 23.22 & 0.7026 & {\color{green_print}0.6667} & {\color{red_print}0.7385} & 31.46 & 0.8723 & {\color{red_print}0.8730} & {\color{green_print}0.8716} \\ 
                            & Fusion   & 5.69  & 0.1953 & {\color{green_print}0.1616} & {\color{red_print}0.2289} & 13.57 & 0.4406 & {\color{green_print}0.4349} & {\color{red_print}0.4463} & 19.34 & 0.6339 & {\color{red_print}0.6352} & {\color{green_print}0.6327} \\ \bottomrule
                            \bottomrule
\end{tabular}%
}
\end{table*}
\vspace{-1em}

\begin{figure}[t]
    \centering
    \begin{subfigure}[b]{0.85\linewidth}
     \centering
        \includegraphics[width=\textwidth, trim={4cm 7cm 2cm 5.5cm}, clip]{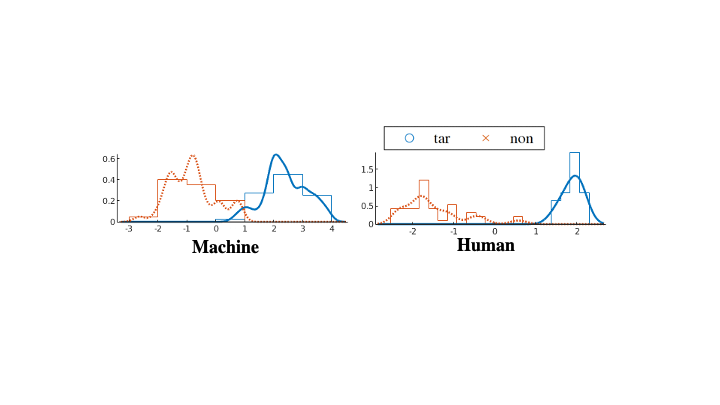}
        \label{fig:llr_spk_a}
      \vspace{-2em}
        \subcaption{}
    \end{subfigure}
    \newline
    \begin{subfigure}[b]{0.9\linewidth}
     \centering
        \includegraphics[width=\textwidth, trim={3cm 7cm 4cm 6cm}, clip]{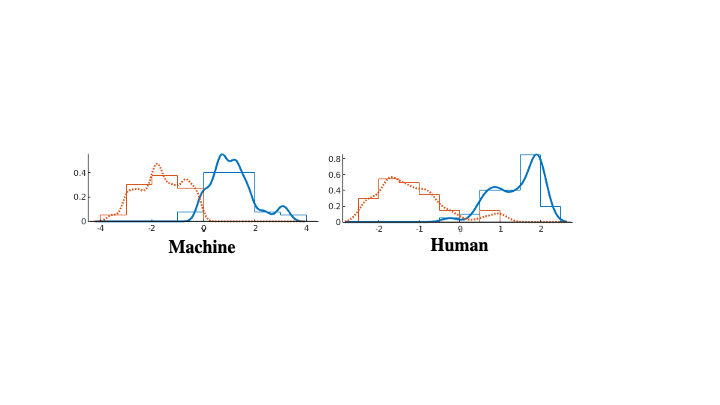}
        \label{fig:llr_spk_b}
       \vspace{-3em}
        \subcaption{}
    \end{subfigure}
        \newline
        \begin{subfigure}[b]{0.9\linewidth}
         \centering
        \includegraphics[width=\textwidth,trim={5cm 7cm 3cm 6cm}, clip]{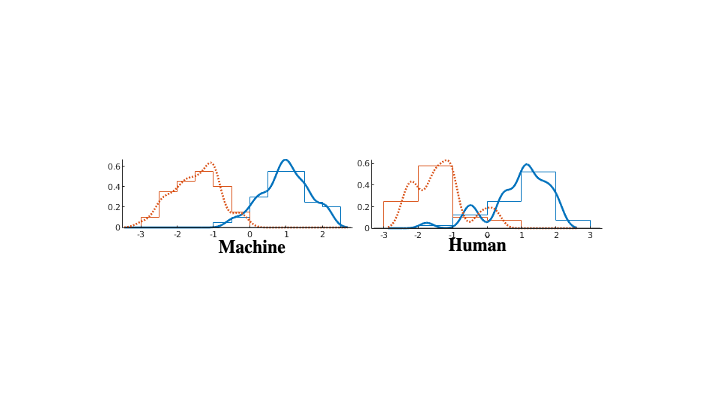}
        \label{fig:llr_spk_c}
      \vspace{-2em}
        \subcaption{}
    \end{subfigure}
    \caption{Histograms of $L^\textrm{tar}$ and $L^\textrm{non}$ per speaker for human and machine responses for trial types: (a) read--read (b) conversation--conversation.  (c) read--conversation.
    }
     \label{fig:llr_spk}
    \vspace{-1em}
\end{figure}
\begin{figure}[t]
    \centering
    \includegraphics[width=0.6\linewidth, trim={0cm, 0cm, 0cm, 1cm},clip]{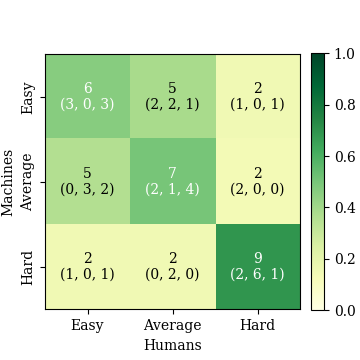}
    \vspace{-1em}
    \caption{Confusion matrix on the difficulty level of speakers for machines versus humans based on the three subsets: easy, average, and hard. Numbers in parentheses represent the count of speakers in the subsets (easy, average, hard) for fused scores.}
    \label{fig:cm}
    \vspace{-1em}
\end{figure}
\section{\label{sec:resndis}Results and Discussion}

\subsection{\label{ssec:performance}Human and Machine Performance}
Table~\ref{tab:perf} presents results for the three speaking-style conditions (read speech -- read speech, conversation -- conversation and read speech -- conversation). Results for native and non-native English listeners are shown separately in Table~\ref{tab:perf} as ``native'' and ``non-native''.  Statistical significance was analyzed using a two-sample Kolmogorov-Smirnoff test~\cite{smirnov_table_1948} and a paired sample McNemar's test~\cite{mcnemar_note_1947} as appropriate. 

The style-matched, read speech -- read speech condition had the best results for both machines and humans in terms of EERs (see Table~\ref{tab:perf}). Although conversation -- conversation trials were also style matched, both humans and machines performed less well than for read speech -- read speech trials, possibly because casual conversations can vary in style depending on the context, topic, and speaker. As hypothesized, the worst performance for both humans and machines was obtained for style mismatched read speech -- conversation trials. Results were consistent with the hypothesis of our previous study~\cite{park_towards_2018} of read and pet-directed speech from the same set of speakers. That study showed that humans consistently performed better than machines in both read speech -- read speech (EER = $19.02$\% versus $30.31$\%) and read speech -- pet-directed speech  (EER = $39.23$\% versus $44.17$\% ) trials. Comparing those results with the present study, we can see that the performance gap between humans and machines decreased with a decreasing amount of style-mismatch (read speech -- conversation). However, caution must be applied, as the EERs could have been affected by score calibration and an improved ASV system used in the present study.

Native listeners performed better than non-natives in all conditions (EERs = $6.96$\% versus $12.39$\% for read speech -- read speech, $15.12$\% versus $23.22$\% for conversation -- conversation, and $20.68$\% versus $31.46$\% for read speech -- conversation, for native versus non-native listeners respectively). All differences were statistically significant ($p<0.05$). It can, therefore, be assumed that, across both style-matched and style-mismatched trials, a listener's fluency in the language being spoken affects perception. Note that in the case of native speakers, humans outperformed machines for style-matched cases (EERs of $6.96$\% versus $14.35$\%; $p<0.05$ and $15.12$\% versus $19.87$\%; $p<0.05$).  In the style-mismatched condition, however, there was no significant difference in the performances between native listeners and machines.  Consistent with~\cite{park_towards_2018, park_target_2019}, native listeners performed better than machines in style-matched trials. There was no significant difference between the performances of non-native listeners and machines on read speech -- read speech and read speech -- conversation trials. But in the case of conversation -- conversation, machines performed better than non-natives (EER of 19.47\% versus 23.22\%, $p<0.05$). 

Finally, the fusion of human and machine scores improved performance significantly in the majority of conditions ($p<0.05$). One exception was read speech -- read speech fusion of native listeners and machine scores ($p=0.05$ for humans versus fusion). The other exception was the conversation -- conversation condition for the fusion of non-native listeners and machine scores ($p=0.053$ for machine versus fusion).  The overall improvement due to fusion was consistent with previous reports~\cite{hautamaki_merging_2013,park_target_2019}, and with our hypothesis that humans and machines use different approaches to speaker discrimination. The small size of the dataset meant that it was not possible to split it into development and evaluation sets while performing fusion, potentially resulting in some over-fitting.

Differences between the same speaker versus different speaker tasks can be studied by comparing $C_\textrm{llr}^\textrm{tar}$ and $C_\textrm{llr}^\textrm{non}$ values in Table~\ref{tab:perf}. The comparison is highlighted in each condition by showing the better value between $C_\textrm{llr}^\textrm{tar}$ and $C_\textrm{llr}^\textrm{non}$ in purple and the worse value in orange. We see that target (same speaker) trials were easier for native listeners when compared to the non-target (different speaker) trials ($p<0.05$). In contrast, for machines the non-target trials were easier ($p<0.05$). On the other hand, non-native listeners were not consistent; they found non-target trials easier in the read speech -- conversation condition and target trials easier in the other two conditions. Recall that $C_\textrm{llr}$ is a measure of soft decisions, and represents the reliability of log-likelihood scores. Table~\ref{tab:perf} shows that native listeners were more reliable than non-natives ($p<0.05$). In subsequent analyses, we only use scores from native listeners.

Female versus male listeners comparison showed that females did better in read speech -- read speech and read speech -- conversation conditions whereas males did better in conversation -- conversation condition (EERs = $9.08$\% versus $10.24$\% for read speech -- read speech, $17.47$\% versus $15.09$\% for conversation -- conversation, and $23.12$\% versus $25.16$\%for read speech -- conversation, for female versus male listeners respectively). All the results were statistically significant ($p<0.05$) and will be studied in detail in follow-up work.


\subsection{\label{ssec:llra}Speaker-level Log-Likelihood-Ratio Analysis}

The histograms of speaker-level log-likelihood-ratios over target trial ($L^\textrm{tar}$) and non-target trial ($L^\textrm{non}$) distributions are shown in Figure~\ref{fig:llr_spk}. For read speech -- read speech trials, the $L^\textrm{tar}$ and $L^\textrm{non}$ distributions are skewed towards correct responses for humans. Increased separation of the target and non-target distributions makes discrimination easier, resulting in better human performance than machine performance (EER $= 6.96$\% versus $14.35$\%). 

In the read speech -- read speech condition, the $L^\textrm{tar}$ distribution for humans showed small variance ($variance=0.05$) and was confined to the positive response region. This was not the case with $L^\textrm{non}$ ($variance=0.57$). Hence, humans were more confident when classifying same voices than different voices, similar to~\cite{park_target_2019}. In contrast, the $L^\textrm{tar}$ distributions had a larger variance and overlapped with $L^\textrm{non}$ distributions in the other two conditions (conversation -- conversation and read speech -- conversation). Hence, there was a decrease in listeners' confidence in classifying both same and different voices.

For machines, variances of $L^\textrm{tar}$ ($variance=0.57$) and $L^\textrm{non}$ ($variance=0.66$) distributions  the read speech -- read speech trials were large and overlapping. This indicates that the degree of uncertainty was higher in the machine scores when compared to humans. There was also a similar overlap in the score distributions for the other two conditions
, which suggests that there may be no particular difference between same speaker and different speaker decisions in machines. 
\subsection{\label{ssec:sl_cllr}Speaker-level Log-Likelihood-Ratio Cost Analysis}
Speaker-level $C_\textrm{llr}$ can be interpreted as the information available to the system (humans or machines) for that speaker. 
The higher $C_\textrm{llr}$ for the speaker, the more confused the system is when distinguishing that speaker from others. 
We divided the speaker population into three subsets for both humans and machine scores, based on the speaker-level $C_\textrm{llr}$. The subset ``\textit{easy}'' (easy to distinguish speakers) comprised thirteen speakers with the lowest $C_\textrm{llr}$ values. Likewise, the ``\textit{hard}'' subset (difficult to distinguish speakers) included the thirteen speakers with the highest $C_\textrm{llr}$ values. The remaining fourteen speakers (neither easy nor hard to distinguish) were classified as ``\textit{average}''.



The confusion matrix in Figure~\ref{fig:cm} provides a visualization of the distribution of speakers in the three subsets. An entry $count_{m_i,h_j}$ denotes the total count of speakers from subset $i$ of machines overlapping with subset $j$ of humans. For example, the second entry (5) is the number of speakers who were \textit{easy} for machines to distinguish but were of average difficulty for humans. A matrix in which entries fall primarily on the diagonal would mean that the difficulty of individual speakers is similar for machines and humans. This is not the case 
in Figure~\ref{fig:cm}. 
Thus, the degree of speaker discrimination difficulty is different for humans and machines. The triplet in parentheses represents the distribution of speakers of the corresponding entry in the matrix into (\textit{easy}, \textit{average}, \textit{hard}) subsets based on the fusion scores. Closer inspection 
shows that diagonal entries get further distributed into off-diagonal entries for the fused scores. The last entry (9) illustrates this point clearly; \textit{hard} speakers for both humans and machines are predominantly \textit{average} (6/9) for the fused system. 

Dissimilarities in the distributions of speakers seem to suggest that humans and machines perform the speaker discrimination tasks differently. 
They might be using different discrimination strategies while performing these 
tasks. The distributions based on the fused scores show a redistribution that is skewed towards correct responses, emphasizing the complementary nature of human and machine scores. 
This highlights instances when machines could assist humans in speaker discrimination.  

\section{\label{sec:con}Conclusion}
The present study was designed to investigate the effects of style variability on speaker discrimination performance for humans and machines. These effects were evaluated using a short-utterance, text-independent speaker discrimination task with read and conversational speech. Experimental results show that both humans and machines performed better in the style-matched condition of read speech -- read speech, followed by conversation -- conversation. This suggests that conversations introduce more variability in the speakers' acoustic spaces. Both humans and machines have the lowest performance for read speech -- conversation, a style-mismatched condition.  Overall, this implies that speaking style variability affects the performances of both humans and machines. Native English-speaking listeners showed higher reliability in performing the tasks than non-natives, and performed better than machines in style-matched conditions. Despite speaking-style variability, listeners were consistently more confident when they are performing the same  speaker tasks versus different speaker tasks, but their overall confidence was the highest for read speech -- read speech trials. 
 Speakers who populate the subsets that humans and machines found easy or difficult to distinguish were not entirely the same. 



\bibliographystyle{IEEEtran}

\bibliography{perception}


\end{document}